\documentclass[a4paper,12pt]{article}

\pdfoutput=1
\usepackage{authblk}
\usepackage{color}
\usepackage{graphicx}
\usepackage{hyperref}
\usepackage{cite}
\usepackage{float}
\usepackage[margin=1cm]{caption}

\usepackage{geometry}
 \newcommand{\nn}{\nonumber}
    
\begin{document}

 \title{Gluon Fusion Contribution to $VHj$ Production at Hadron Colliders}
\author[1]{Pankaj Agrawal}
\author[2]{Ambresh Shivaji}
\affil[1]{Institute of Physics, Sachivalaya Marg, 
          Bhubaneswar, Odisha, INDIA - 751005}
\affil[2]{Regional Centre for Accelerator-based Particle Physics, 
          Harish-Chandra Research Institute,  
          Chhatnag Road, Jhusi, Allahabad 211019, India}

%  \author{
% Pankaj Agrawal and Ambresh Shivaji 
%        }
 
  \maketitle

 \begin{abstract}
 \noindent
 We study the associated production of an electroweak vector boson and the Higgs boson 
  with a jet via gluon-gluon fusion. At the leading order, these processes 
 occur at one-loop level. The amplitudes 
 of these one-loop processes are gauge invariant and finite. Therefore, their 
 contributions towards the corresponding hadronic cross sections and 
 kinematic distributions can be calculated separately. We present results for 
 the Large Hadron Collider and its discussed upgrades. We find that the gluon-gluon
 one-loop process gives dominant contribution to the $\gamma H j$ production. 
 We observe a destructive interference effect in the $gg\to Z H j$ amplitude. 
 We also find that in the high transverse momentum and central rapidity region, 
 the $ZHj$ production cross section via gluon-gluon fusion becomes 
 comparable to the cross section 
 contributions coming from quark-quark and quark-gluon channels.  

 \end{abstract}
{\tt PACS numbers:} 12.38-t, 12.38.Bx, 13.85.Lg \\
{\tt Keywords:} LHC, Gluon Fusion, Higgs, One-loop
 
  \vfill
\begin{flushright}
%TU-xxx\\
% HRI-P-13-09-002\\
RECAPP-HRI-2014-020
\end{flushright}

%  \section{Introduction}
 \newpage

    With the discovery of the Higgs boson \cite{Aad:2012tfa,Chatrchyan:2012ufa} 
    at the Large Hadron Collider
  (LHC), the standard model and its symmetry-breaking mechanism have been validated.  
  The CMS and ATLAS collaborations are measuring its properties.The Higgs boson
  was discovered through $g g \to H$ production mechanism. Since then signals
  of other production mechanisms have also been examined. In particular, the
  associated production of the Higgs boson with an electroweak boson has been 
  explored \cite{cmshw, cmshwth}. To ensure that the discovered Higgs boson 
  is indeed the standard model Higgs boson, there is a need to detect as many of 
  its signatures as possible. Furthermore, since there is no signal that points 
  to new physics beyond the standard model, there is a need to look for 
  standard model processes that do not have large but accessible cross sections, 
  and can be enhanced/modified by new physics effects.
  
   The LHC and its proposed upgrades provide us an opportunity to explore
  two types of processes in more details: a) the processes which,
 in the standard model, begin at the one-loop level (one-loop being the leading order (LO));
 b) gluon-gluon scattering processes. As the centre-of-mass energy increases,
 the gluon-gluon luminosity increases, making many more processes observable. 
 Study of such relatively rare processes  
 is complementary to new physics searches at high energy colliders.
% Although the 
% existence of the standard model Higgs boson appears to be a reality~\cite{Aad:2012tfa,Chatrchyan:2012ufa}, 
% there is no conclusive evidence of new physics signals in the available LHC data yet. Since
% the effects of new physics are expected to be small, we also need to carefully study 
% all those processes for which the standard model predictions are rather small.
% At hadron colliders scattering processes can be initiated by quark-quark (QQ), 
% quark-gluon (QG) and gluon-gluon (GG) channels. 
% It is well known that at hadron colliders such as the Large Hadron Collider (LHC), the gluon luminosity 
 %increases with increasing collider energy. The GG initiated processes, therefore, 
 %can have sizable cross sections. 
 Thus, the LHC provides a unique opportunity for testing many 
 of the standard model predictions which was not possible at earlier high energy colliders.
 For example, many one-loop gluon-gluon fusion processes are/will be  accessible only at 
 the LHC~\cite{Agrawal:2012df,deFlorian:1999tp,Agrawal:1998ch,Campbell:2014gua,Melia:2012zg,Campanario:2012bh}. 
 These gluon fusion one-loop processes can be studied both in the signal and 
 background categories. 
 
%  \section{Structure of the Amplitude}
 
\begin{figure}
\begin{center}
 \includegraphics[width=0.6\textwidth]{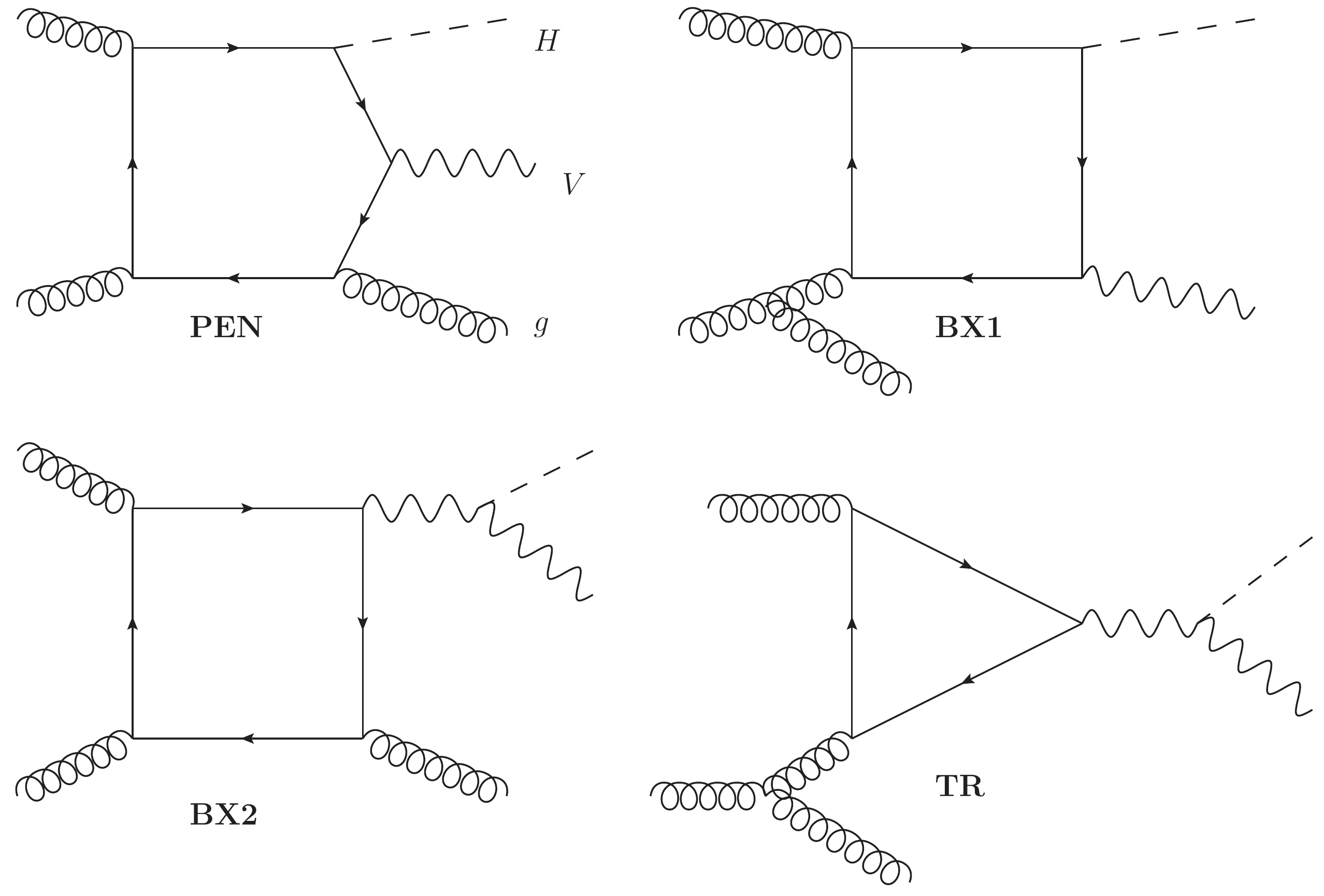}
  \caption{Classes of diagrams contributing to the $gg \to VHj$ processes. The complete set of diagrams can be 
  generated by permuting the external legs. The leading order $gg \to \gamma Hj$ amplitude
  receives contribution only from the pentagon diagrams.}
\label{fig:VHj}
\end{center}
 \end{figure}

 In this letter, we are interested in the gluon-gluon contribution to the 
 $p p \to V~ H~ j+X$ hadronic processes, i.e., 
%  /reduce
  \begin{equation}
  g + g \to V~ H~  j, \nn
  \end{equation}
 where $V$ is an electroweak vector boson and `{\em j}' stands for a light-quark, or
 gluon initiated jet. The amplitude of the process $gg \to W^{\pm}Hj$ is 
 trivially zero due to the electromagnetic charge conservation. Therefore, $V$ would refer to a 
 photon or a $Z$ boson. This process occurs at the one-loop via triangle, box, and pentagon
 diagrams. Since quarks carry both the electroweak and color charges, the 
 leading order contribution comes from quark-loop diagrams.
 The prototype quark-loop diagrams are shown in Fig.~\ref{fig:VHj}~\cite{Binosi:2008ig}. 
%These figures are drawn using the  {\tt JaxoDraw} package~\cite{Binosi:2008ig}. 
(We have not displayed some of the triangle and 
 bubble diagrams which do not contribute due to the vanishing color factor.)
  For the $\gamma Hj$ case, only pentagon class of diagrams contribute. The box diagrams 
 with $qqH$ coupling do not contribute because of the Furry's theorem.
 Due to the same reason, the leading order 
 $gg \to \gamma H$ amplitude vanishes. There are a total 24 pentagon diagrams. However, due to charge 
 conjugation symmetry only 12 of those are independent. 
 The full amplitude is proportional to the symmetric color factor $d^{abc}$. 
%  Symbolically, 
%    \begin{eqnarray}
%  {\cal M}(gg \to H\gamma j) = i  \frac{d^{abc}}{2} {\cal M}^{\rm PEN}_{\rm V}.
%   \end{eqnarray}
 
 In the $ZHj$ case, box and triangle diagrams also contribute. 
Once again there are 12 independent pentagon (PEN) diagrams.
There are two types 
 of box diagrams depending on the nature of the Higgs coupling. 
  The number of 
 independent box diagrams with $ZZH$ coupling (BX2) is 3. Because of 
 the nonsymmetric color factors, in the case of PEN and BX2 amplitudes
 both the vector and axial-vector pieces of the $Z$ boson coupling with quarks 
 contribute.  
 In the case of box diagrams with $qqH$ coupling (BX1), there are 9 ($3\times3$) 
 independent diagrams due to self gluon coupling.There are also 3 independent triangle 
 (TR) diagrams.
 In the case of BX1 and TR diagrams, contribution of the vector part of the $qqZ$ coupling
 vanishes due to the Furry's theorem. The vector and axial-vector parts 
 of the full amplitude are proportional to the symmetric color factor $d^{abc}$ and 
 antisymmetric color factor $f^{abc}$ respectively.
 Except the top quark, all other quarks in the loop are treated massless. Therefore, in the diagrams 
 involving $qqH$ coupling, we take only the top quark contribution. Since the axial-vector 
 part of the amplitude is proportional to the isospin quantum number $T_3^q$, only third 
 generation quarks (which have large mass difference) give contribution.
 The triangle and box diagrams involving $ZZH$ coupling are anomalous,
therefore both the top and bottom quark contributions are taken. We ignore the 
contribution of the top quark loop of the BX2 diagrams in the case of vector
$ttZ$ coupling due to its large mass
and no compensating factor from the couplings.

 The amplitudes of the gluon fusion processes under consideration have following structure:
 
  \begin{eqnarray}
   {\cal M}(gg \to \gamma H j) &=& i  \frac{d^{abc}}{2} {\cal M}^{\rm PEN}_{\rm V}, \\
{\cal M}(gg \to Z H j) &=& i \frac{d^{abc}}{2} {\cal M}_{\rm V} + \frac{f^{abc}}{2} {\cal M}_{\rm AV}, \\
{\cal M}_{\rm V} &=& {\cal M}^{\rm PEN}_{\rm V} - {\cal M}^{\rm BX2}_{\rm V}, \\
{\cal M}_{\rm AV} &=& {\cal M}^{\rm PEN}_{\rm AV} + {\cal M}^{\rm BX1}_{\rm AV} - {\cal M}^{\rm BX2}_{\rm AV} - {\cal M}^{\rm TR}_{\rm AV},
 \end{eqnarray}
where the subscripts `V' and `AV' stand for the vector and axial-vector parts of the amplitude. 
% The superscripts 
% `PEN',`BX1',`BX2' and `TR' refer to the pentagon diagrams, box diagrams with $ttH$ coupling, box diagrams with $ZZH$ 
% coupling and triangle diagrams respectively. 
Since the vector and axial-vector parts of the amplitude have symmetric 
and antisymmetric color factors, they do not interfere at the amplitude-squared level.
 Note that the leading order gluon fusion process $gg \to VHj$ contributes to the corresponding hadronic process at 
 next-to-next-to-leading order (NNLO) in strong coupling parameter $\alpha_s$. 
 It also contributes to the QCD  next-to-leading order 
 (NLO) correction of $gg \to VH$~\cite{Dicus:1988yh,Kniehl:1990iva,Altenkamp:2012sx}.

   To calculate this amplitude, we follow a semi-numerical approach. As a first step,  
 the traces of gamma-matrices in the prototype diagrams are calculated using {\tt FORM}~\cite{Vermaseren:2000nd} in 
 $n$ dimensions. This demands an implementation of a suitable $n-$dimensional prescription for 
 $\gamma^5$ in {\tt FORM}. The gamma-matrices traces involving $\gamma^5$ can also be calculated in four 
 dimensions. This will lead to spurious/anomalous terms in amplitudes suffering from chiral 
 anomaly. It is known that such anomalous terms contribute to the quark mass independent rational 
 piece of the amplitude~\cite{Shivaji:2011ww,Shivaji:2013cca}. Since the standard model is free 
 from anomaly, the spurious terms are expected to get canceled between the up-type and down-type 
 quark contributions of a given quark generation. Since all the gamma-matrices in the trace are 
 contracted either with the momenta or polarizations, the trace calculation does not lead to any 
 explicit $n$ dependence. The amplitude at this stage can be cast in terms of tensor integrals. It has 
 rank-four, five-point functions as the most complicated tensor integrals. In standard notations, a 
 rank-four, five-point tensor integral can be written as,
 \begin{equation}
  E^{\mu\nu\rho\sigma} = \int \frac{d^nl}{(2\pi)^n} \frac{l^\mu l^\nu l^\rho l^\sigma}{d_0d_1d_2d_3d_4},
 \end{equation}
 where $l$ is the loop momentum, $n$ is the space-time dimension and $d_i$'s are loop momentum
 dependent scalar propagators. 
 The tensor integrals are known to be expressible in terms of scalar integrals~\cite{Passarino:1978jh}. 
 This is the most important part of any multileg one-loop computation. We reduce pentagon tensor 
 integrals into box tensor and scalar integrals using the fact that in four dimensions the 
 loop momentum ($l$) can be expressed as a linear combination of four independent external momenta which 
 are available in a five-point ($2 \to 3$) process. The same fact is used to write a 
 pentagon scalar integral in terms of five box scalar integrals~\cite{Shivaji:2013cca,vanNeerven:1983vr}. 
 Since the pentagon tensor integrals 
 are ultraviolet (UV) finite, their reduction into box tensor and scalar integrals can be carried out 
 in four dimensions consistently~\cite{Bern:1993kr,Binoth:2005ff}.
 We follow the methods of Oldenborgh and Vermaseren (OV) \cite{vanOldenborgh:1989wn} to reduce the box and 
 lower-point tensor integrals.  This part of the reduction is performed in $n=4-2\epsilon$ dimensions 
 to regulate the bad ultraviolet behaviour of tensor integrals. In OV method, the organization of tensor integrals 
 in terms of generalized Kronecker deltas promises a greater numerical stability of the reduction procedure.
 The complete one-loop tensor reduction library, {\tt OVReduce}, is developed (in {\tt FORTRAN}) to carry out the 
 numerical reduction of one-loop tensor integrals required in $2 \to 3$ processes~\cite{Agrawal:1998ch}.
 The singular structure of any one-loop amplitude is, thus, encoded in various scalar integrals of box, triangle, 
 bubble and tadpole types. 
 We use {\tt OneLOop} library~\cite{vanHameren:2010cp} to compute all the required scalar integrals. 
 It uses dimensional regularization to regulate both the UV and infrared (IR) singularities.
We require only one prototype pentagon and BX2 amplitudes to generate the vector part 
of the full amplitude. On the other hand, to generate the full axial-vector part of the amplitude, 
we require three prototype pentagon amplitudes, three BX1 amplitudes, one BX2 amplitude and
one TR amplitudes. This is to ensure that 
 permutation of momenta and polarizations is consistent with the permutation of external
 legs and their couplings with the massive quarks. 
 
We perform many checks to ensure the correctness of the $VHj$ amplitudes.
The amplitudes are expected to be both ultraviolet and infrared 
(due to massless quarks) finite. This is an important check on the amplitudes. 
The IR finiteness holds for each quark loop diagram~\cite{Shivaji:2010aq}. 
 The UV finiteness is expected only in gauge invariant subamplitudes, however, 
there are amplitudes which are individually UV finite from naive power counting. For example, 
each pentagon amplitude and each axial-vector part of BX1, BX2 and TR amplitudes are UV finite.
As an ultimate check, we have checked the gauge invariance of
amplitudes with respect to all the gauge currents. We do it
numerically by replacing their polarization  vectors with their respective
4-momenta, $\epsilon_\mu(k) \to k_\mu$. The vector part of the amplitude 
is gauge invariant with respect to the $Z$ boson. However, due to the 
explicit breaking of the chiral symmetry in presence of massive quarks, 
the axial-vector part of the amplitude is not gauge invariant with respect 
to the $Z$ boson. We have checked explicitly that the quantum anomaly is 
canceled between the top and bottom quark contributions. 
The three-body phase space is generated using the phase space generator
 {\tt RAMBO}~\cite{Kleiss:1985gy}. 
Because of a very large and complicated expression of the amplitudes,
we calculate the (helicity/ polarized) amplitudes before squaring them. The parton 
level phase space integration and convolution of partonic cross section with the gluon 
distribution functions are done using the monte-carlo integration method based on {\tt VEGAS}~\cite{Vegas:1980}.
 To minimize the computation time, we perform the evaluation in a parallel environment with the help 
 of the {\tt AMCI} package~\cite{Veseli:1997hr} based on the {\tt PVM}~\cite{PVM:1994} parallel networking 
 software. The computation 
 of multileg, one-loop amplitudes often suffers from the issue of numerical instability due to the small 
 Gram determinants for certain phase space points. Since the number of such phase space points is not very 
 large, we systematically ignore their contribution by implementing a set of Ward-Identities. More details on 
 this can be found in Refs.~\cite{Campanario:2011cs,Agrawal:2012as}.

%  \section{Results}
 
      \begin{table}[h]
%    \begin{table}[H]
 \begin{center}
  \begin{tabular}{|c|c|c|c|c|c|c|}
   \hline
   $\sqrt{\rm S}\;(\rm TeV)$& $8$& $13$& $14$&  $33$& $100$ \\
   \hline
   \hline
   & & & & &\\
      $\sigma^{\gamma Hj,\; \rm LO}_{\rm GG}\;[\rm ab]$ & 72 & 268 & 320 & 2029 & 13435 (12734) \\
   & & & & &\\
%    \hline
   $\sigma^{\gamma Hj,\; \rm LO}_{\rm QQ+QG}\;[\rm ab]$ & 26 & 81 & 96 & 491 & 2667 (1312) \\
   & & & & &\\
   \hline
  \end{tabular}
 \end{center}
%  \vspace{0.5cm}
\caption{A comparison of parton channels contributing to $pp \to \gamma H j$ 
hadronic cross section.The 
numbers in bracket in 100 TeV column are with a minimum $p_T^j$ cut of 50 GeV.}
\label{table:XS-gph}
\end{table} 
 
\begin{figure}[t]
 \includegraphics[width=0.5\linewidth]{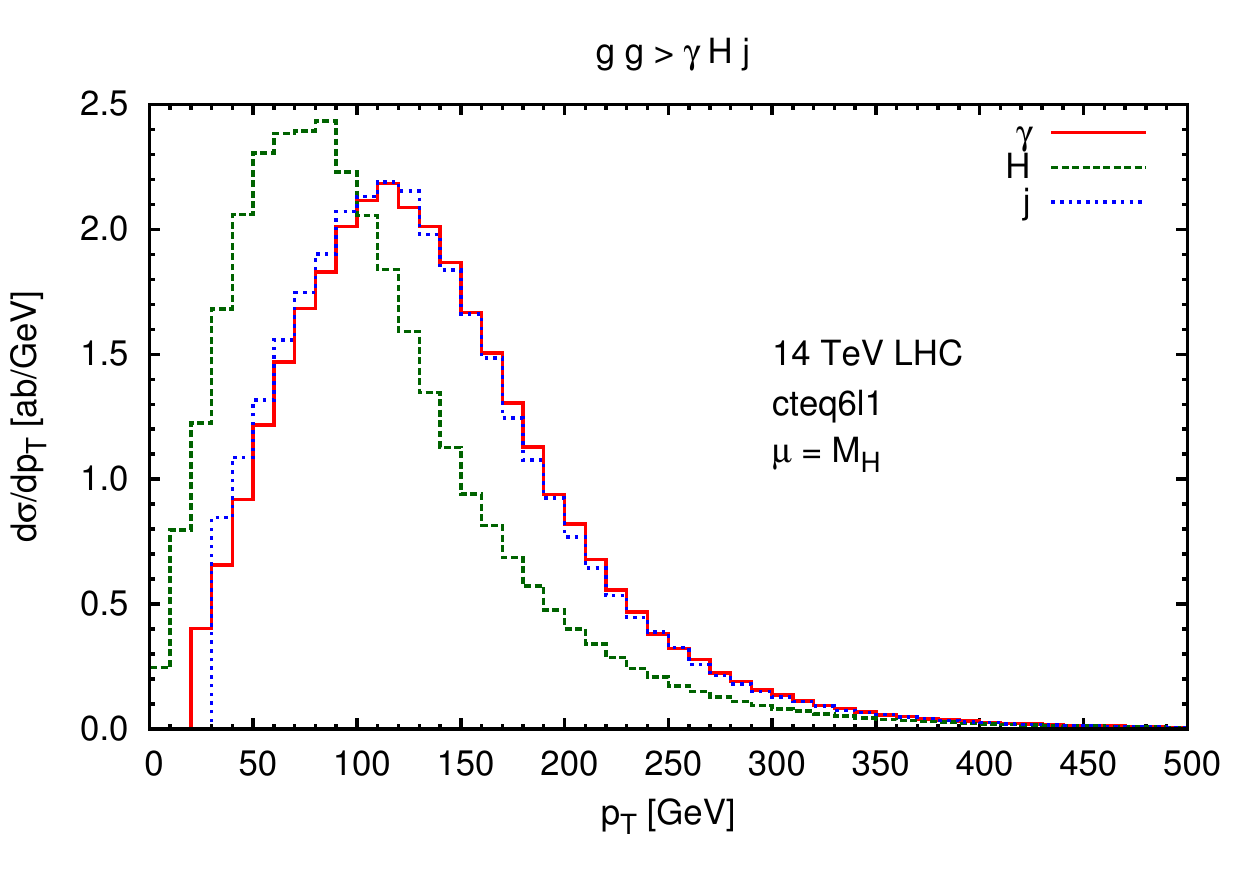}
  \includegraphics[width=0.5\linewidth]{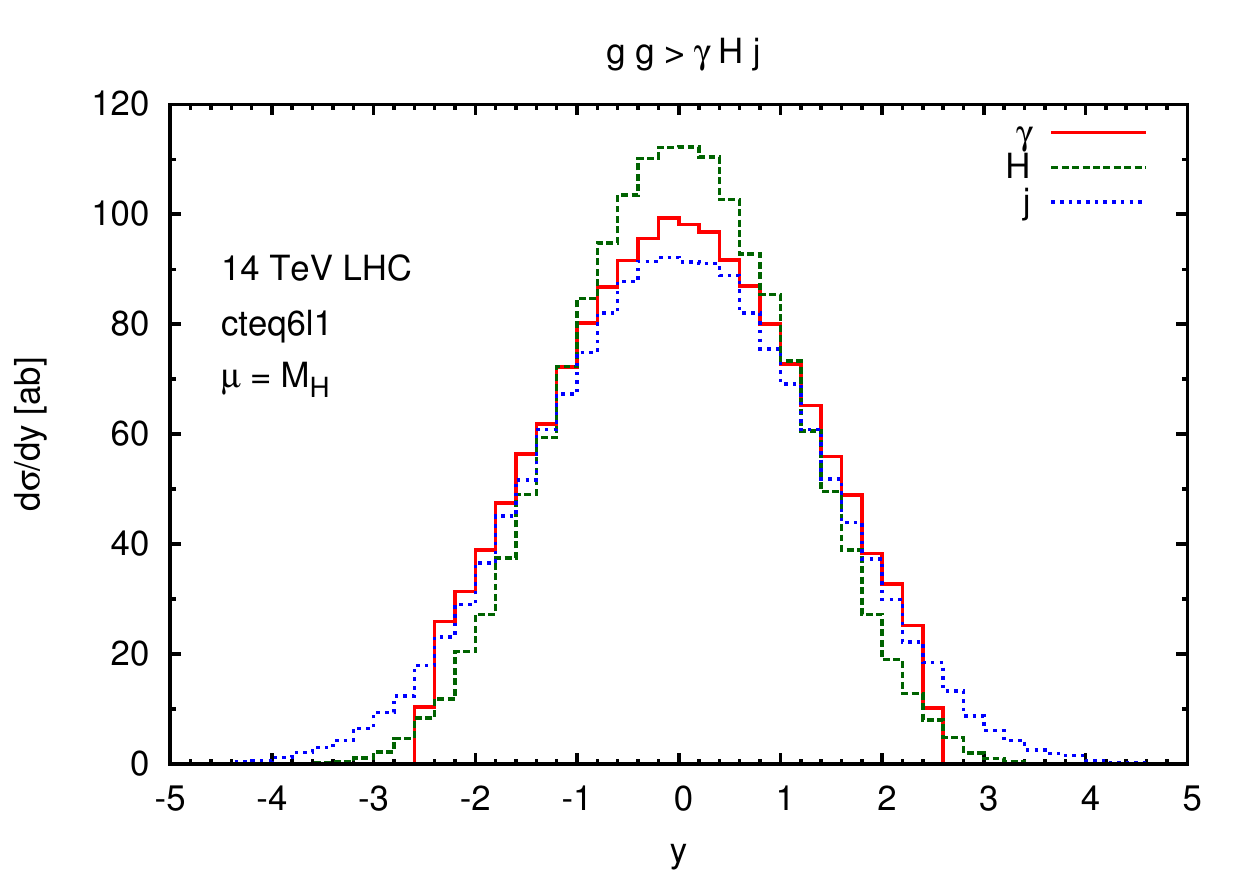}
  \caption{Transverse momentum and rapidity distributions of the final state particles in $gg \to \gamma H j$.}
  \label{fig:plots-gph}
 \end{figure}

 We now present the numerical results for the $V H j$ processes. 
 We have taken $m_t = 173\; {\rm GeV}$ and $M_H = 126 \; {\rm GeV}$. Following kinematic 
 cuts are applied to obtain the results discussed below, 
 \begin{equation}
     p_T^j > 30\; {\rm GeV},\; p_T^\gamma > 20\; {\rm GeV},\; |y^j| < 4.5,\; |y^\gamma| < 2.5,\;
     \Delta R^{\gamma j} > 0.4. \nn
 \end{equation}
 We use {\tt cteq6l1} parton distribution functions~\cite{Nadolsky:2008zw} and choose $\mu_R = \mu_F = M_H$ as the common 
 central scale for renormalization and factorization. 
 In Table~\ref{table:XS-gph}, we compare the LO gluon-gluon (GG) contribution to $pp\to \gamma Hj$ with the contribution 
 from the leading order quark-quark (QQ) and quark-gluon (QG) initiated process at various collider centre-of-mass energies. 
 The contribution from QQ+QG channel involving $\gamma\gamma H$ and $Z\gamma H$ one-loop couplings is expected to be 
 negligible because of the large electroweak coupling suppression.
 The leading order QQ+QG channel contribution is calculated using {\tt MG5}~\cite{Alwall:2011uj}.
%We use {\tt cteq6l1} parton distributions and the scales as $\mu_F = \mu_R = M_H$. 
 Due to small bottom quark mass and
  low bottom quark flux in proton, the QQ+QG contribution is smaller than the GG contribution. However, 
 the hadronic cross section in the gluon-gluon fusion channel itself is very small ($\sim $ 0.3 fb at 14 TeV) due to the 
 top quark propagator suppression in the pentagon amplitudes. With this cross section, in run II of the LHC, we would
 expect about 100 such events. We also see that the ratio of GG channel and QQ+QG channel
 cross sections increases as the centre-of-mass energy increases. It is expected due to the 
 faster increase in the GG luminosity with centre-of-mass energy. 
 With proper background suppression strategies, one may be able to observe 
 this process at HL-LHC and HE-LHC.
In figure~\ref{fig:plots-gph}, we display the transverse momentum ($p_T$) and rapidity 
 ($y$) distributions of the final state particles in the $\gamma Hj$ case. Note that the $p_T$ distributions for both the photon 
 and the jet are well defined in the collinear region ($p_T \to 0$). This is so because the amplitude for 
 $gg \to \gamma H j$ receives contribution only from massive quark loop diagrams\footnote{ A collinear singular configuration 
 involves three massless lines at a vertex.}. 

\begin{figure}
   \begin{center}
  \includegraphics[width=0.6\linewidth]{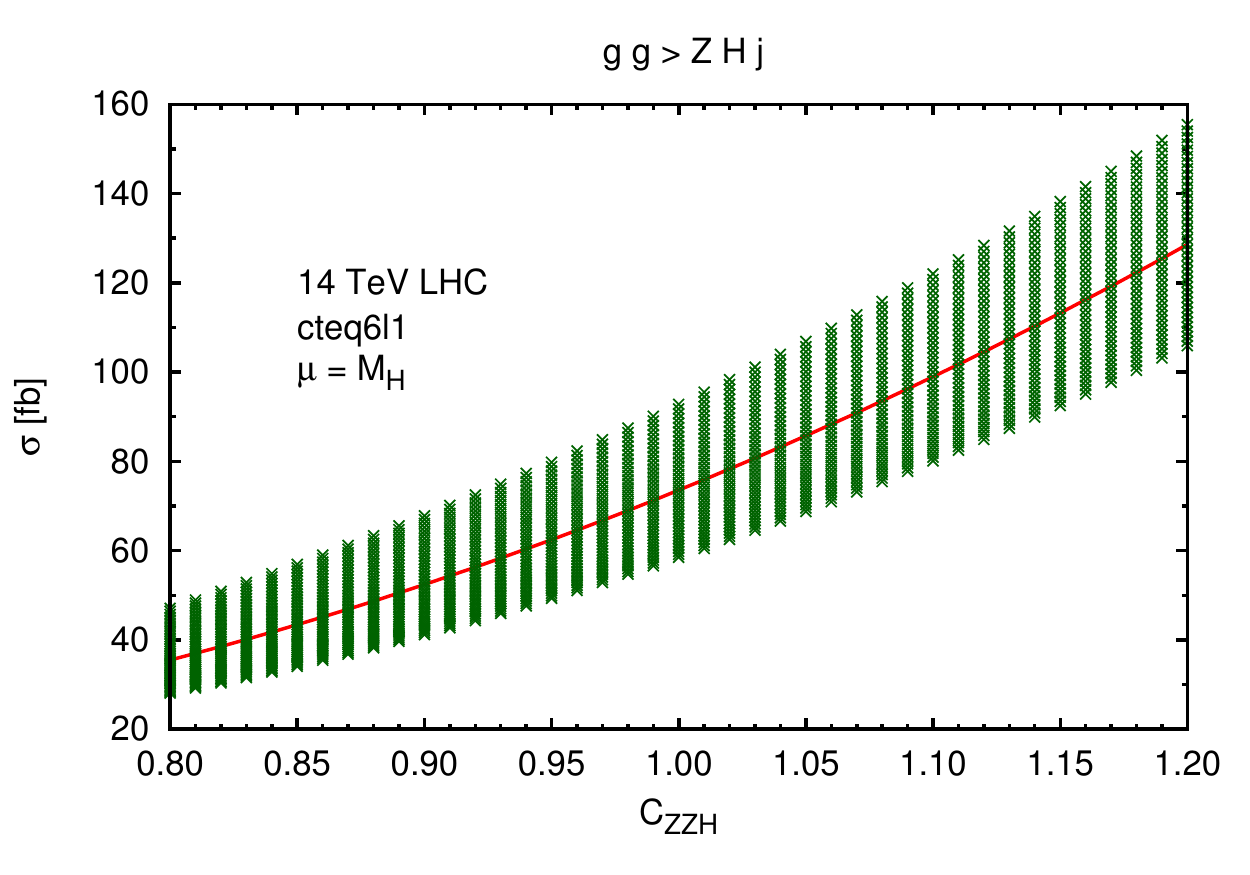}
     \end{center}
   \caption{Variation of $ZHj$ cross section with the anomalous parameters $C_{\rm ZZH}$ and $C_{\rm ttH}$
   at $\sqrt{\rm S} =$ 14 TeV. The central line corresponds to $C_{\rm ttH} =1$, while the vertical spread 
   about this line is due to the variation of $C_{\rm ttH}$ between 0.80 and 1.20, $C_{\rm ttH} =1.20$ being 
   at the lower end.}
             \label{fig:scan-anml}
  \end{figure}

We have seen that unlike in the case of $\gamma Hj$, the $ZHj$ production via gluon-gluon fusion also involves box 
and triangle diagrams with both massive and massless quarks in the loop. 
At 14 TeV, the $ZHj$ cross section is about 73 (64) fb with {\tt cteq6l1} ({\tt cteq6m}) 
parton distributions and fixed scale choice $\mu=M_H$. 
We observe that almost all the contribution to the cross 
section comes from the axial-vector or color antisymmetric part of the amplitude. At 14 TeV LHC, the 
vector and axial-vector contributions are,
% \begin{equation}
 $\sigma_{\rm V} = 0.10\;{\rm fb} \;\; {\rm and} \;\; \sigma_{\rm AV} = 73.36\;{\rm fb} $
% \end{equation}
respectively.
 \begin{figure}[t]
%   \begin{center}
  \includegraphics[width=0.5\linewidth]{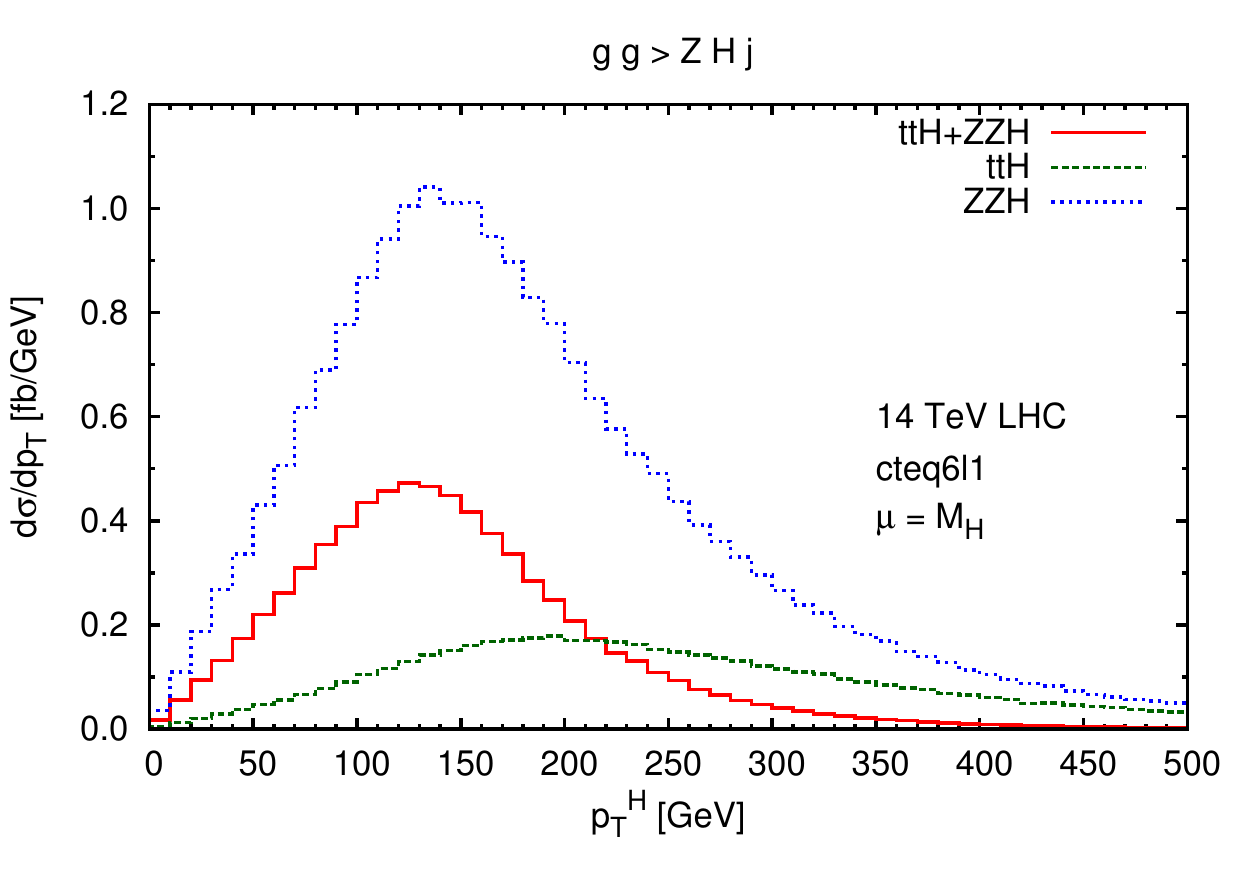}
  \includegraphics[width=0.5\linewidth]{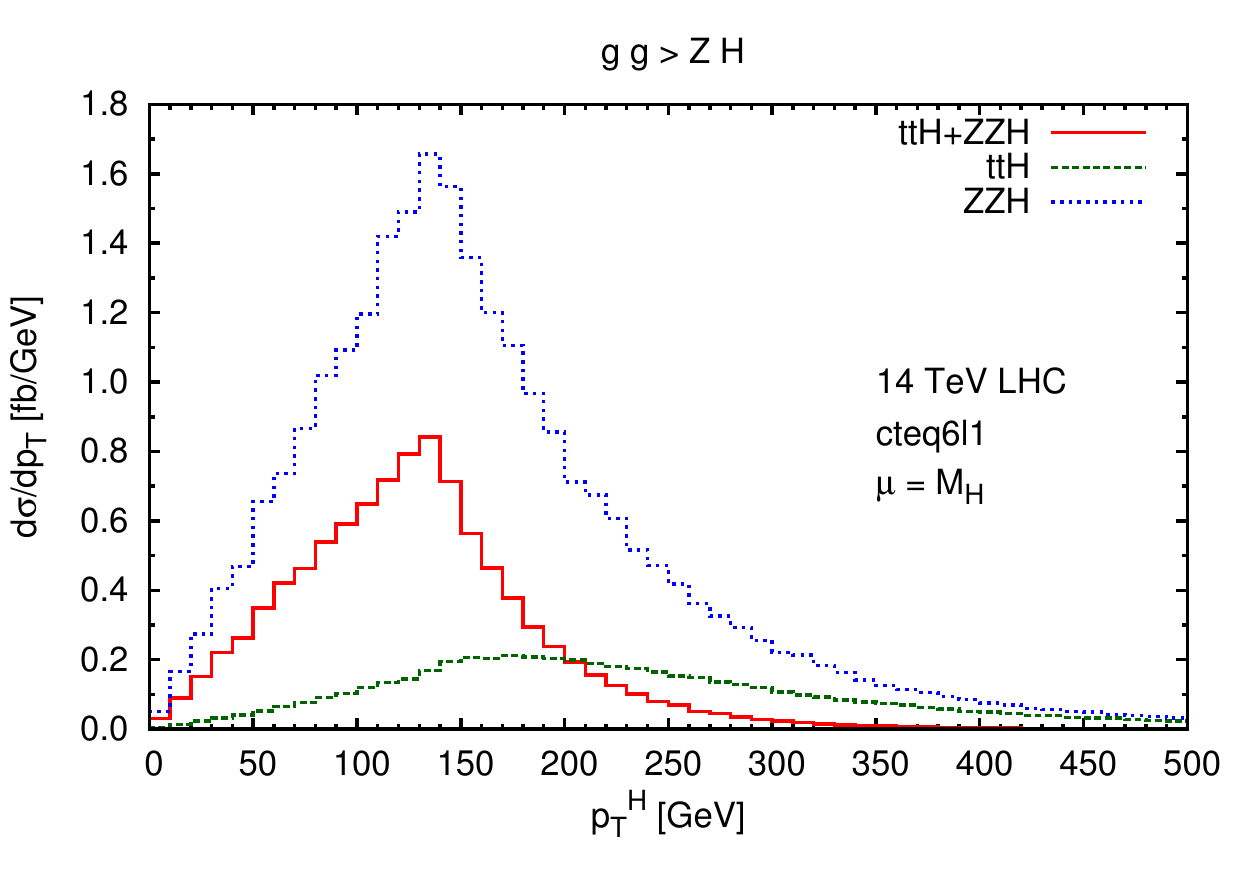}
   \caption{$p_T$ distributions of the Higgs boson displaying the strong interference effect between the gauge invariant 
             subamplitudes involving $ttH$ and $ZZH$ couplings in $gg \to ZHj$ and $gg\to ZH$ processes.}
             \label{fig:pth-interference}
%              \end{center}
  \end{figure}
There are many gauge invariant 
subamplitudes in the $gg\to ZHj$ amplitude. The nature of interference among them can be considered as an important 
prediction of the standard model. We have two sets of gauge invariant subamplitudes -- one 
involving $ttH$ coupling and the other involving $ZZH$ coupling. Like in many associated Higgs production
processes, we find that there is a very strong destructive interference between the two sets 
of amplitudes~\cite{Agrawal:2012ga,Nishiwaki:2013cma}. To illustrate this feature, in figure~\ref{fig:pth-interference}, 
we have given $p_T$ distributions of the Higgs boson considering only one gauge invariant set of amplitudes at a time.
 This is very similar to the destructive interference effect seen in $gg \to ZH$ amplitude; it is 
also shown in the right panel of figure~\ref{fig:pth-interference}.
To quantify the interference effect in the $ZHj$ production, we have calculated the contributions of different gauge invariant sets towards 
the hadronic cross section. At 14 TeV, the cross sections are, 
% \begin{equation}
$ \sigma_{\rm ttH} =52.56\;{\rm fb} ,\;\; \sigma_{\rm ZZH} = 211.78\;{\rm fb} ,\;\; {\rm and }\;\; \sigma_{\rm ttH + ZZH} = 73.56 \;{\rm fb}.$
% \end{equation}
% 
In presence of new physics in the Higgs sector, such a strong interference effect may lead to a very different predictions for both 
the cross section and kinematic distributions. For example, if we introduce deviations (due to new physics) in the $ttH$ and $ZZH$ 
couplings through the scale factors $C_{\rm ttH}$ and $C_{\rm ZZH}$ respectively, the $ZHj$ cross section at 14 TeV can be parametrized 
as, 
\begin{equation}
 \sigma_{\rm GG}^{pp \to ZHj}|_{\sqrt{\rm S} = 14 {\rm TeV}} = 52.56\; C_{\rm ttH}^2 + 211.78\; C_{\rm ZZH}^2 -190.78\; C_{\rm ttH}\; C_{\rm ZZH}.
\end{equation}
The standard model prediction corresponds to $C_{\rm ttH} = C_{\rm ZZH} = 1$. In figure~\ref{fig:scan-anml}, 
we give the variation of this cross section by changing the anomalous parameters by 20 $\%$ around their standard model values. 
Note that the effects of these parameters on the cross section is opposite in nature within the range they 
are varied. Also, because of a larger contribution from $ZZH$ coupling diagrams which is due to the
inclusion of light quark diagrams, the net cross section is more 
sensitive to variation in $C_{\rm ZZH}$.

  \begin{figure}
 \includegraphics[width=0.5\linewidth]{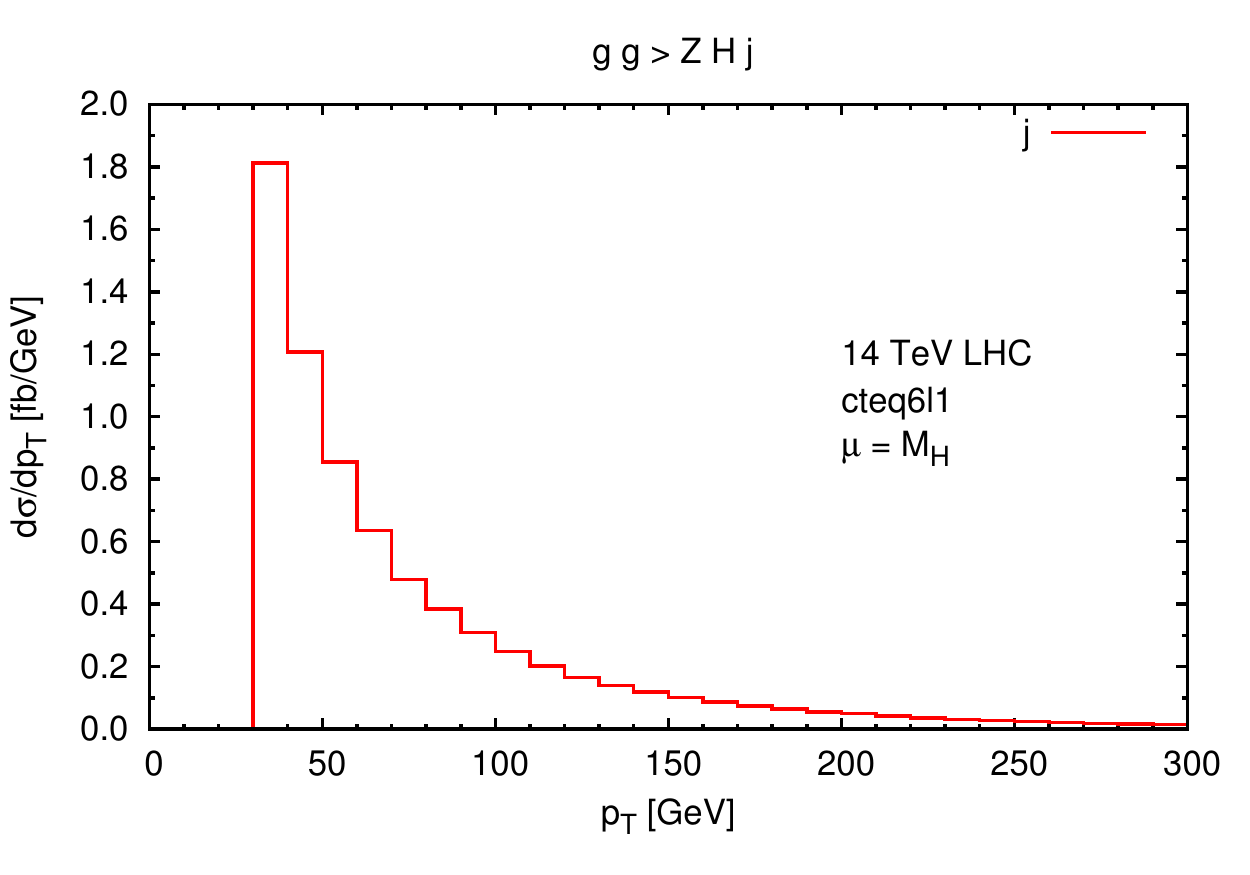}
  \includegraphics[width=0.5\linewidth]{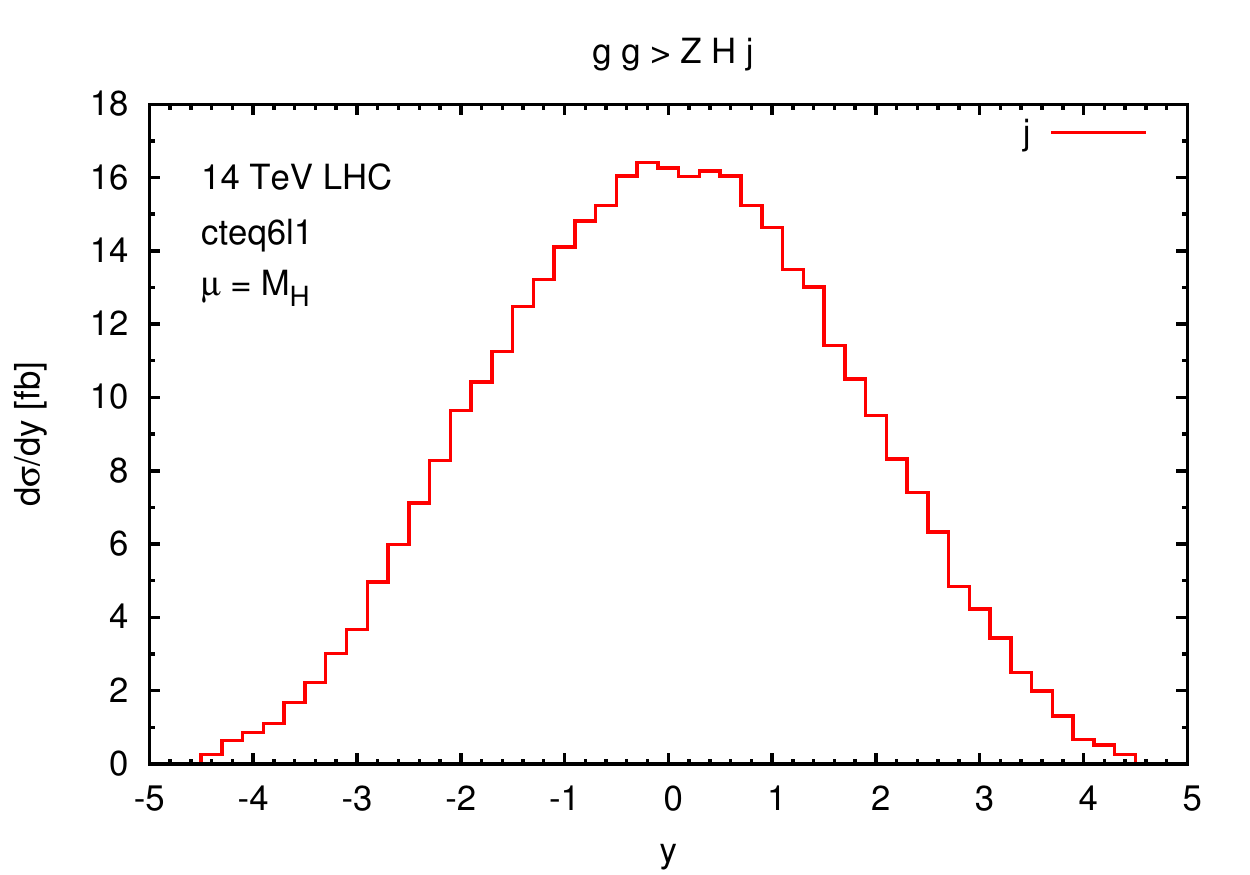}
  \caption{Transverse momentum and rapidity distributions of the jet in $gg \to Z H j$.}
  \label{fig:distj-gzh}
 \end{figure}

     \begin{table}[t]
 \begin{center}
  \begin{tabular}{|c|c|c|c|c|c|c|}
   \hline
   $\sqrt{\rm S}\;(\rm TeV)$& $8$& $13$& $14$&  $33$& $100$ \\
   \hline
   \hline
   & & & & &\\
     $\sigma^{ZHj,\;\rm LO}_{\rm GG}\;~[\rm fb]$ 
     & 14.96 & 63.93 & $73.56^{+52.6\%}_{-32.1\%}$ & 535.30 & 4270.71 (2788.90) \\  
   & & & & &\\
    $\sigma^{ZHj,\;\rm LO}_{\rm QQ+QG}\;~[\rm fb]$ 
    & 95.12 & 226.80 & $256.52^{+11.7\%}_{-9.9\%}$ & 955.00 & 4143.34 (2715.82) \\
   & & & & &\\
%    \hline
     $\sigma^{ZHj,\;\rm NLO}_{\rm QQ+QG}\;[\rm fb]$ 
     & 114.20 & 266.90 & $302.60^{+3.0\%}_{-3.4\%}$ & 1076.00 & 4414.00 (3087.21)  \\
   & & & & &\\
     $R_{1} = \frac{\sigma^{ZHj,\rm LO}_{\rm QQ+QG}}{\sigma^{ZHj,\rm LO}_{\rm GG}}$ 
     & 6.36 & 3.55 & 3.49 & 1.78 & 0.97 (0.97)  \\
   & & & & &\\
     $R_{2} = \frac{\sigma^{ZHj,\rm NLO}_{\rm QQ+QG}-\sigma^{ZHj,\rm LO}_{\rm QQ+QG}}{\sigma^{ZHj,\rm LO}_{\rm GG}}$ 
     & 1.27 & 0.63 & 0.63 & 0.23 & 0.06 (0.13)  \\
   & & & & &\\
    \hline
     & & & & &\\
      $\sigma^{ZH,\;\rm LO}_{\rm GG}\;~~[\rm fb]$ 
      & 24.60 & 82.38 & $97.70^{+32.1\%}_{-22.7\%}$ & 569.90 & 3764.63 \\
   & & & & &\\
   \hline
  \end{tabular}
 \end{center}
%  \vspace{0.5cm}
 \caption{ A comparison of partonic channels contributing to $pp \to Z H j$ 
hadronic cross section. We use {\tt cteq6l1} ({\tt cteq6m}) for LO (NLO) results and $\mu_F = \mu_R = M_H$.
 The numbers in bracket in 100 TeV column are with a minimum $p_T^j$ cut of 50 GeV.}
 \label{table:XS-gzh}
\end{table}
% % 
   \begin{figure}[ht!]
%      \begin{center}
   \includegraphics[width=0.5\linewidth]{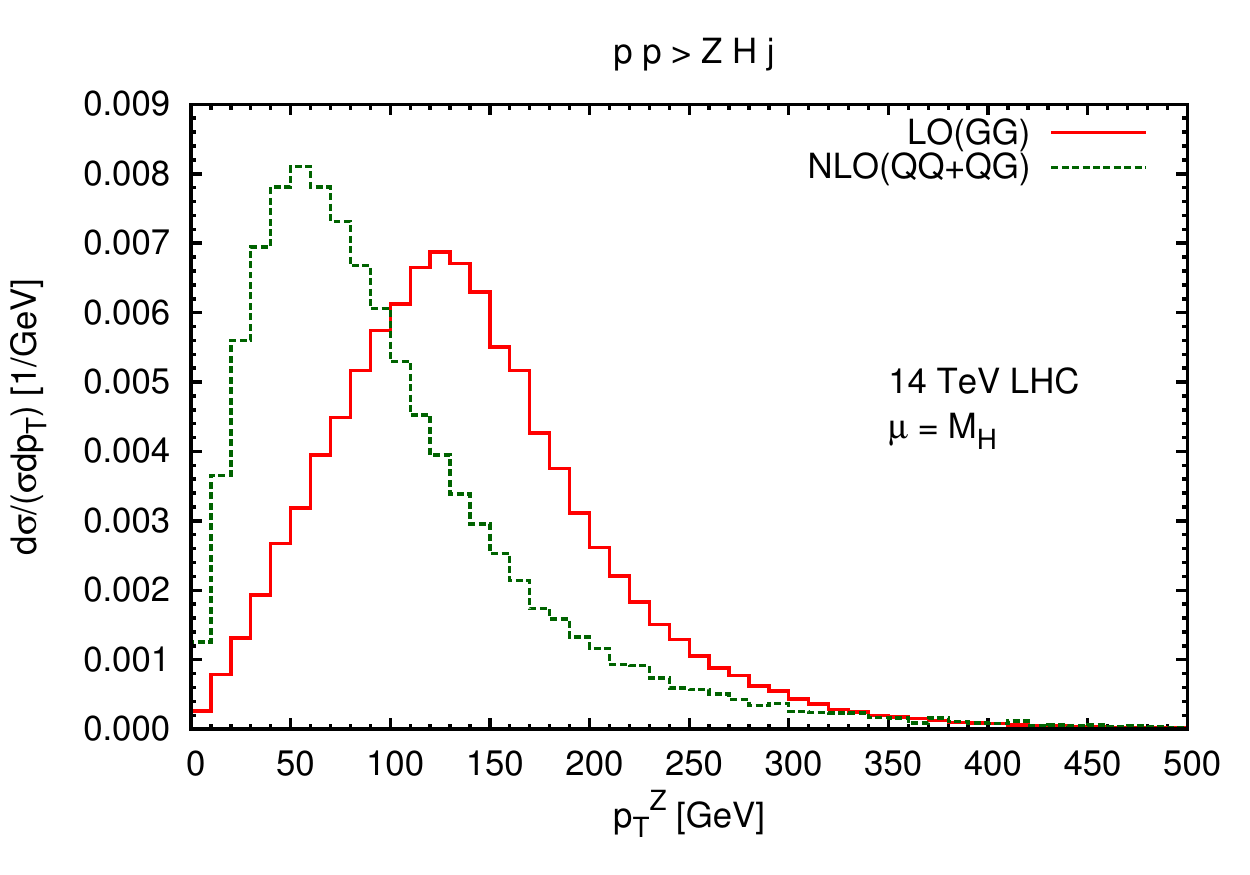}
   \includegraphics[width=0.5\linewidth]{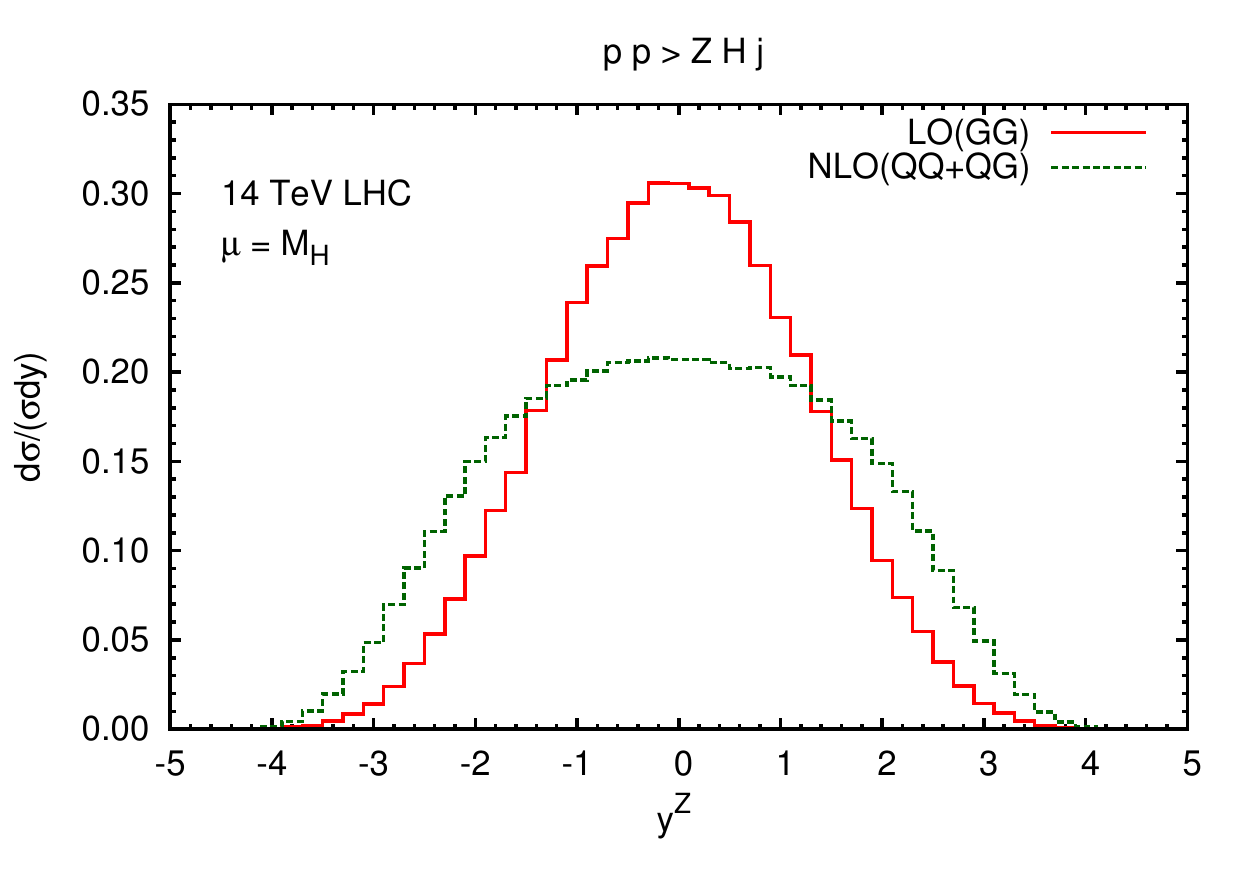}
  \caption{A comparison of the LO (GG) and NLO (QQ+QG) contributions to the transverse momentum and rapidity 
            distributions (normalized by cross sections)
             of the $Z$ boson. }
            \label{fig:ptz-gg-vs-qq}
%               \end{center}
 \end{figure}
 
In figure~\ref{fig:distj-gzh}, the $p_T$ and rapidity distributions for the jet is shown. Since the final state gluon 
can originate as radiation from the initial state gluons (as in BX1 and TR diagrams), the majority of events come from 
low $p_T$ region of the jet. We have calculated the $gg \to ZHj$ hadronic cross section at 8, 13, 14, 33 and 100 TeV collider
centre-of-mass energies. In going from 8 TeV to 14 TeV, the cross section increases by about a factor of 5 due to an 
increase in the gluon flux at a higher energy. In Table~\ref{table:XS-gzh}, we have compared the gluon-gluon (GG)
 contribution to $pp \to ZHj$ with the LO and NLO quark-quark (QQ) and quark-gluon (QG) predictions at 
various collider centre-of-mass energies.The NLO correction to this process has been computed in recent past~\cite{JiJuan:2012tk,Luisoni:2013cuh}. 
We have obtained the NLO results using {\tt MG5\_aMC@NLO}~\cite{Alwall:2014hca} 
with {\tt cteq6m} parton distributions. As expected, the relative 
importance of the GG contribution increases with the increasing collider energy. To illustrate this, 
we have displayed two ratios $R_1(=\sigma^{ZHj,\;\rm LO}_{\rm QQ+QG}/\sigma^{ZHj,\;\rm LO}_{\rm GG})$ and 
$R_2(= (\sigma^{ZHj,\;\rm NLO}_{\rm QQ+QG} - \sigma^{ZHj,\;\rm LO}_{\rm QQ+QG})/\sigma^{ZHj,\;\rm LO}_{\rm GG})$ in the table.
% we have given the ratio $R_{1}$, which is $\sigma^{ZHj,\;\rm LO}_{\rm QQ+QG}/\sigma^{ZHj,\;\rm LO}_{\rm GG}$
% We also display a ratio $R_{2}$, $(\sigma^{ZHj,\;\rm NLO}_{\rm QQ+QG} - \sigma^{ZHj,\;\rm LO}_{\rm QQ+QG})/\sigma^{ZHj,\;\rm LO}_{\rm GG}$. 
The ratio $R_2$ suggests that even at 8 TeV, the GG contribution becomes comparable to the NLO {\em enhancement}
of QQ+QG contribution.  Nevertheless, the net contribution from the 
QQ+QG channel remains dominant at all energies. In the 14 TeV column, the numbers are presented with the percentage 
scale uncertainties. For that we have varied the scale by a factor of 2 about the central value $\mu =M_H$. A large 
scale uncertainty in the GG predictions is typical to many other gluon-gluon fusion processes which can be reduced by 
including the higher order corrections. One example of such a process is $ g g \to H$ which has been widely studied in 
the literature~\cite{Dittmaier:2011ti}.
The GG numbers for 100 TeV should be interpreted carefully. 
In the last row of the table, we also report $gg\to ZH$ hadronic cross sections. Note that, the $gg \to ZHj$ 
cross section exceeds $gg \to ZH$ cross section at $\sqrt{\rm S}=$100 TeV for a minimum $p_T^j$ cut of 
30 GeV. This is a reflection of the fact that at such 
high energies, the cross section is also influenced by the large-log terms like ${\rm ln}(p_T^2/{\rm S})$. Due to this, 
the fixed-order perturbative cross section prediction cannot be reliable below certain $p_T$
and one needs to resum these large logarithms. Therefore, 
in the table, we have also reported the cross section at 100 TeV with $p_T^j > 50$ GeV.

We have also compared the normalized $p_T$ and rapidity distributions of the $Z$ boson at LO(GG) and NLO(QQ+QG). It is shown in 
figure~\ref{fig:ptz-gg-vs-qq}. The difference between the two contributions is consistent with the 
fact that at higher $p_T$, the region of parton distributions with smaller momentum fraction $x$ is easily accessible and 
gluon distributions are more important in this region. 
This comparison suggests that at high $p_T$ and low rapidity the GG contribution is relatively more important along with 
the dominant QQ+QG channel contribution.
The $p_T$ and rapidity distributions of the Higgs also show a similar characteristic 
difference, however, the same is not true for the distributions of the jet.

%  \section{Conclusion}

In conclusion, we have studied the gluon-gluon fusion contribution to the $pp \to V H j $ 
process, where $V \in \{\gamma, Z \}$. It is a NNLO contribution in $\alpha_s$. We find that 
the hadronic cross section for $\gamma H j$ is only about 0.3 fb at 14 TeV LHC, 
however, it is much larger than the leading order quark-quark and quark-gluon
contribution which is only about 0.01 fb at 14 TeV. This raises the possibility
of observing the production of the Higgs boson in association with a photon
with sufficient integrated luminosity at the LHC. In the $ZHj$ case, apart
from the GG fusion contribution, we have used {\tt MG5\_aMC@NLO} to estimate
the NLO corrections to the QQ+QG contribution. We find that at 13 or 14 TeV
centre of mass energy, the 
gluon-gluon contribution is comparable to the NLO corrections. At higher centre-of-mass
energies, the GG contribution becomes even more important and can become
comparable to the QQ+QG contributions.
In the $ZHj$ case, we also observe a 
destructive interference between the two gauge invariant sets of amplitudes 
involving $ttH$ and $ZZH$ couplings. Any modification to these couplings due to 
new physics effects can spoil the interference effect and lead to a very different
prediction. A suitable $p_T$ and rapidity cuts can be employed to enhance the signal
from this channel while comparing it with quark-quark and quark-gluon channels. For realistic predictions 
the parton shower effects should also be included which is beyond the scope of the present work.
Like many other gluon-gluon fusion LO processes, the processes considered here too 
suffer from a large scale uncertainty. By computing higher order QCD 
corrections, the scale uncertainty can be reduced. However, this will be quite 
challenging at the NLO as it requires evaluation of one-loop six-point functions 
and two-loop five-point functions.

 \section*{Acknowledgements}
 We would like to acknowledge the use of Cluster Computing facility available 
 at the Institute of Physics, Bhubaneswar. AS would like to acknowledge the 
 hospitality of the institute where the present work was initiated.
 AS would also like to thank Manoj Mandal, Marco Zaro and Rikkert Frederix for their 
 technical help with the {\tt MG5\_aMC@NLO} package.
The work of AS is partially supported by funding available 
from the Department of Atomic Energy, Government of India, 
for the Regional Centre for Accelerator-based
Particle Physics, Harish-Chandra Research Institute.

\end{document}